\PassOptionsToPackage{table}{xcolor}
\documentclass[sigplan,nonacm,10pt]{acmart}
\acmConference[PLDI'22]{ACM SIGPLAN Conference on Programming Language Design and Implementation}{June 13--17, 2022}{San Diego, CA, USA}
\acmYear{2022}
\acmISBN{} 
\acmDOI{} 
\startPage{1}

\setcopyright{none}

\usepackage{xspace}

\makeatletter
\newcommand{\AnonyNoteFirst}[0]{%
  \if@ACM@anonymous%
  \footnote{Name changed to preserve anonymity.\label{anonynote}}%
  \else\fi%
}
\newcommand{\AnonyNote}[0]{%
\if@ACM@anonymous%
\footref{anonynote}%
\else%
\fi%
}
\newcommand{\Quickstrom}[0]{%
  \if@ACM@anonymous%
  \textsc{Bl\aa{}haj}%
  \else%
  Quickstrom%
  \fi\xspace}
\newcommand{\Specstrom}[0]{%
  \if@ACM@anonymous%
  \textsc{Billy}%
  \else%
  Specstrom%
  \fi\xspace}
\newcommand{\QuickLTL}[0]{%
    \if@ACM@anonymous%
    \textsc{Kallax}%
    \else%
    QuickLTL%
    \fi\xspace}
\makeatother
\usepackage{mathpartir}
\usepackage{mathtools}
\usepackage{combelow}   
\usepackage{booktabs}   
\usepackage{subcaption} 
\usepackage{tikz}

\usepackage{siunitx}
\usepackage{pgfplots}
\pgfplotsset{width=7cm,compat=1.3}
    
\usetikzlibrary{shadows}
\usepackage{cleveref}
\newcommand{\SNext}{\operatorname{%
  \tikz[baseline]{
    \draw[line width=.12ex]
      (0,.6ex) circle (.8ex);
      \draw[line width=.12ex] (-0.8ex,-0.4ex) -- (0.8ex,-0.4ex);
  }}}{}
\newcommand{\WNext}{\operatorname{%
  \tikz[baseline]{
    \draw[line width=.12ex]
      (0,.6ex) circle (.8ex);
      \draw[line width=.12ex] (-0.8ex,1.6ex) -- (0.8ex,1.6ex);
  }}}{}
\newcommand{\DNext}{\operatorname{%
  \tikz[baseline]{
    \draw[line width=.12ex]
      (0,.6ex) circle (.8ex);
    \draw[line width=.12ex]
      (0,.6ex) circle (.2ex);
  }}}{}
\newcommand{\Next}{\operatorname{%
  \tikz[baseline]{
    \draw[line width=.12ex]
      (0,.6ex) circle (.8ex);
  }}}{}
  \newcommand{\Eventually}{\operatorname{%
  \tikz[baseline]{
    \draw[line width=.12ex,line join=round]
      (0ex,.6ex) -- (.95ex,1.55ex) -- (1.9ex,.6ex) -- (.95ex,-.35ex) -- cycle;
  }}}{}
  \newcommand{\Always}{\operatorname{%
  \tikz[baseline]{
    \draw[draw=white] (-0.2ex,0ex) -- (1.7ex,0ex);
    \draw[line width=.12ex,line join=round]
      (0ex,-.2ex) -- (0ex,1.3ex) -- (1.5ex,1.3ex) -- (1.5ex,.-.2ex) -- cycle;
  }}}{}
  \newcommand{\Until}{\mathbin{%
  \mathcal{U}}}{}
  \newcommand{\Release}{\mathbin{%
  \mathcal{R}}}{}
\begin{document}

\title{\Quickstrom: Property-based Acceptance Testing with LTL Specifications}         


\author{Liam O'Connor}
\orcid{0000-0003-2765-4269}             
\affiliation{
  \institution{University of Edinburgh}            
  \city{Edinburgh}
  \country{Scotland}                    
}
\email{l.oconnor@ed.ac.uk}          

\author{Oskar Wickstr\"om}
\orcid{0000-0003-1789-6926}             
\affiliation{
  \institution{Monoid Consulting}           
  \city{Malm\"o}
  \country{Sweden}                   
}
\email{oskar@wickstrom.tech}         

\begin{abstract}
We present \Quickstrom\AnonyNoteFirst, a property-based testing system for acceptance testing of interactive applications. Using \Quickstrom, programmers can specify the behaviour of web applications
as properties in our testing-oriented dialect of Linear Temporal Logic (LTL) called \QuickLTL\AnonyNote, and then automatically test their application against the given
specification with hundreds of automatically generated interactions. \QuickLTL extends existing finite variants of LTL for the testing use-case, determining likely outcomes from partial traces whose minimum length is
itself determined by the LTL formula. This temporal logic is embedded in our specification language,
\Specstrom\AnonyNote, which is designed to be approachable to web programmers, expressive for writing specifications, and easy to analyse.
Because \Quickstrom tests only user-facing behaviour, it is agnostic to the implementation language of the system under test.
We therefore formally specify and test many implementations of the popular TodoMVC benchmark, used for evaluation and comparison across various web frontend frameworks and languages.
Our tests uncovered bugs in almost half of the available implementations.
\end{abstract}




\maketitle

\section{Introduction}
Property-based testing, such as that of QuickCheck~\citep{qc}, is a popular bug testing
methodology whereby software is specified in the form of logical
properties, and automatically tested against randomly-generated inputs
to find possible counterexamples to those specifications. Property-based testing specifications
are more high-level than unit tests, and facilitate greater maintainability with less effort.
Unlike unit testing, property-based testing allows the programmer to specify
the behaviour of a module without also specifying the expected behaviour of the module's user, i.e.\ the expected inputs to a function.

With the increasing use of web browser technology for user interfaces of
 applications, automatic testing of these interfaces using browser testing technology such as Selenium WebDriver~\citep{selenium},
 has become more necessary. To write a test in Selenium, the programmer must first script a specific interaction with their application's user interface,
 and then test that the interaction produces the expected result. For example, to test the property:
 \newcommand*\keystroke[1]{%
  \tikz[baseline={([yshift=0.2em]key.south)}]
    \node[%
      draw=black,
      fill=white,
      drop shadow={shadow xshift=0.25ex,shadow yshift=-0.25ex,fill=black,opacity=0.75},
      rectangle,
      rounded corners=2pt,
      inner sep=1pt,
      line width=0.5pt,
      font=\scriptsize\sffamily
    ](key) {#1\strut}
  ;
}
 \begin{center}
  \emph{When I click \keystroke{$\ \mathsf{Cancel}\ $}, I should return to the main menu.}
 \end{center}
 The programmer would write a script that first simulates a click to the \keystroke{$\ \mathsf{Cancel}\ $} button and then
 inspects the state of the user interface to confirm that we have indeed returned to the main menu. Other properties, however,
 are not so simple, such as this invariant:
 \begin{center}
  \emph{I should not reach the finances page without logging in.}
 \end{center}
Or this \emph{temporal} property:
\begin{center}
\emph{The menu should never be disabled forever.}
\end{center}
 These properties cannot be easily translated into a Selenium script, because Selenium tests, like unit tests, require the programmer to
 specify not just the intended behaviour of the application but also the expected behaviour of the application's user. This is where \Quickstrom comes in.

 \Quickstrom~\citep{quickstrom_deanony} is an in-development open-source tool which uses
 property-based testing techniques to enable automatic behavioural acceptance testing of web user interfaces from high-level specifications.
 Using a simple specification language, engineers inform \Quickstrom of their desired properties, as well as how to interact with their user interface.
 Then, \Quickstrom generates and tests hundreds of possible interactions, just as property-based testing libraries generate inputs, checking that the given properties
 are not violated.

In conventional property-based testing frameworks, the properties that make up specifications  usually take the form of equations relating inputs to expected outputs of functions under test.
\Quickstrom, however, is not designed for testing functions, but for testing \emph{whole applications}. These applications cannot be viewed as functions---instead they are \emph{reactive} systems: they continuously respond to
signals such as user actions and environmental events.

One of the most common logics used to specify reactive systems is Linear Temporal Logic (LTL)~\citep{ltl}, a logic equipped with
temporal modalities to describe \emph{behaviours}: completed, infinite traces of a system's execution. Our tests, however, only produce finite traces: as only a finite number of actions can be
taken, only a finite prefix of a desirable behaviour can be observed.
 Our dialect of LTL, called \QuickLTL, is extended to accommodate this testing use-case. It is a \emph{multi-valued} version of LTL defined for finite, partial traces
 whose minimum length is determined by the given formula. The logic is multi-valued to enable \Quickstrom to give \emph{presumptive} answers for when the formula
 cannot be definitively proven nor refuted by the steps taken so far.
 The syntax and semantics of \QuickLTL are given in \Cref{sec:quickltl}.

 \QuickLTL is embedded in our bespoke specification language \Specstrom. This language is designed to be familiar to web programmers, expressive for writing specifications, and simple to analyse.
 In addition to writing \QuickLTL formulae, engineers also use \Specstrom to tell \Quickstrom which actions to take and which events to expect when running tests. Details of the design of \Specstrom and examples of its
 use are given in \Cref{sec:specstrom}.

 Our framework is designed for \emph{acceptance testing}, that is, it only tests the user-observable behaviour of the application as a whole. Therefore, \Quickstrom specifications are independent of the
 language used to implement the application under test. TodoMVC is a widely-implemented benchmark and sample application for a variety of web application frameworks and languages. We have converted the (informal) English
 specification of TodoMVC to a formal \Specstrom specification, and used \Quickstrom to test its various implementations, uncovering bugs and problems in more than a third of the available implementations. Our specification and our test results are discussed in \Cref{sec:todomvc}.

 \subsection*{Contributions}
 \begin{itemize}
\item The design and implementation of the \Quickstrom tool itself, including its \Specstrom interpreter and its test executor based on Selenium WebDriver,
\item The \QuickLTL temporal logic, a multi-valued dialect of Linear Temporal Logic for partial traces which incorporates minimum constraints on the
length of the trace. We specify its semantics by formula progression and provide examples of its use.
\item The design and implementation of the specification language \Specstrom, which includes a variety of features, such as control over evaluation, which make it easy to
specify systems. We specify an egg timer as a worked example.
\item A formal specification of the TodoMVC benchmark in \Specstrom/\QuickLTL, and our evaluation of various implementations of this benchmark against our formal specification, in which
we find faults in over one third of available implementations.
 \end{itemize}
 \section{LTL and \QuickLTL}\label{sec:quickltl}

 \begin{figure}
  \[
  \begin{array}{lclcll}
\multicolumn{3}{l}{\text{Formulae:}}\\
 \varphi, \psi & ::=  & \multicolumn{3}{l}{ p\ \mid\ \neg \varphi\ \mid\ \top\ \mid\ \bot} \\
    & \mid & \varphi \land \psi &\mid & \varphi \lor \psi \\
  & \mid & \Next{\varphi} & & & \textit{(next)} \\
            & \mid & \Always{\varphi} &\mid& \Eventually{\varphi}  & \textit{(henceforth/eventually)} \\
            & \mid & \varphi \Until \psi  & \mid & \varphi \Release \psi  & \textit{(until/release)} \\
  p  & \in & \multicolumn{3}{l}{\Sigma \rightarrow \{ \top, \bot \} } & \text{predicates}\\
\sigma & \in & \Sigma &&& \text{states}\\
\rho & \in & \Sigma^\omega &&& \text{behaviours}\\
  \end{array}
  \]
  \caption{Syntax of LTL}
  \label{fig:pnueli_ltl_syntax}
  \end{figure}

\begin{figure}
  \[
  \begin{array}{lcl}
    \multicolumn{3}{l}{\text{For}\ \rho = \sigma_0\sigma_1\sigma_2\dots:}\\
    \rho \models p & \Leftrightarrow &p(\sigma_0)\\
    \rho \models \varphi \land \psi  &\Leftrightarrow& \rho \models \varphi\ \text{and}\  \rho \models \psi \\
    \rho \models \neg \varphi &\Leftrightarrow& \rho \not\models \varphi \\
    \rho \models \Next \varphi &\Leftrightarrow& \sigma_1\sigma_2\dots \models \varphi \\
    \rho \models \Eventually \varphi &\Leftrightarrow& \text{There exists an $i$ such that}\ \sigma_i\dots \models \varphi \\
    \rho \models \Always \varphi &\Leftrightarrow& \text{For all $i \geq 0$, }\ \sigma_i\dots \models \varphi \\
    \rho \models \varphi \Until \psi &\Leftrightarrow& \text{There exists an $i$ such that}\ \sigma_i\dots \models \psi \\
    & & \text{and for all}\ j < i\text{,}\ \sigma_j\dots \models \varphi\\
    \rho \models \varphi \Release \psi &\Leftrightarrow& \text{For all $i \geq 0$,}\ \sigma_i\dots \models \psi\ \text{or} \\
    & & \text{there exists}\ j < i\text{ such that}\ \sigma_j\dots \models \varphi
  \end{array}
  \]
  \caption{Semantics of LTL}
  \label{fig:pnueli_ltl}
  \end{figure}

 Linear Temporal Logic~\citep{ltl} is a modal logic that describes \emph{behaviours}: infinite, linear sequences of states ordered by time. The syntax of
 LTL is given in \Cref{fig:pnueli_ltl_syntax} and its semantics in \Cref{fig:pnueli_ltl}. When our behaviours are the completed traces or executions of our application,
 we can use LTL to write its specification. For instance, we can express invariants using the modality $\Always$ (read ``henceforth'' or ``always''), as in this invariant which states that users should not be able to access the ``Finances'' page without being logged in:
 $$\Always (\mathsf{LoggedIn} \lor \mathsf{page} \neq \texttt{"Finances"})$$
 Invariants are an example of \emph{safety properties}, which say that ``bad'' states will not be reached. It is straightforward to find counterexamples to
 safety properties by testing, as safety properties are exactly those that can be refuted in a finite number of steps~\citep{alpernschneider},
 but many specifications also include
 \emph{liveness properties}, which say that a ``good'' state will (eventually) be reached. We can express liveness properties by using the modality $\Eventually$ (read ``eventually''), dual to $\Always$. For example, this property
 states that a menu will eventually be enabled:
 $$\Eventually \textsf{menuEnabled}$$
 Counterexamples to
 liveness properties are more difficult to find via testing, as they take the form of infinite traces where the desired ``good'' state is never reached---no finite amount of
 testing will ever produce a complete counterexample. Conversely, if, rather than search for counterexamples, we instead search for a positive witness that the property holds,
 liveness properties become easy and safety properties become hard.

 We can combine $\Eventually$ with $\Always$ to state that the menu will be enabled infinitely often; or, equivalently,
that the menu will never be disabled forever:
$$\Always \Eventually \textsf{menuEnabled}\qquad\neg\Eventually\Always\textsf{menuDisabled}$$
Both of the temporal operators $\Eventually$ and $\Always$ are special-cases of the more general temporal operators $\Until$ (read ``until'') and its dual $\Release$ (read ``release'') respectively, as can be
seen in identities 6--7 of \Cref{fig:identities}. Using these operators we can express more sophisticated requirements on the
ordering of events, such as these (equivalent) properties that state that we cannot access a secret page without logging in first:
$$ \mathsf{LogIn}\Release\neg\mathsf{SecretPage}\quad\quad \neg(\neg \mathsf{LogIn} \Until \mathsf{SecretPage})$$
All of these operators can be thought of as fixed points of expansion identities involving the $\Next$ (read ``next'') operator, such
as identities 8--11 of \Cref{fig:identities}. We can also use $\Next$ in our specifications, such as this example that describes a flashing
screen, alternating between dark and light:
$$ \Always (\mathsf{dark}\land \Next \mathsf{light} \lor \mathsf{light}\land \Next \mathsf{dark})$$

    \begin{figure}
      \begin{eqnarray}
      \neg \Eventually \varphi & = & \Always \neg \varphi  \\
      \neg \Always \varphi & = & \Eventually \neg \varphi  \\
      \neg \Next \varphi & = & \Next \neg \varphi  \\
      \neg (\varphi \Until \psi) & = & \neg \varphi \Release \neg \psi\\
      \neg (\varphi \Release \psi) & = & \neg \varphi \Until \neg \psi\\[0.5em]
      \Eventually \varphi & = & \top \Until \varphi \\
      \Always \varphi & = & \bot \Release \varphi \\[0.5em]
      \Always \varphi & = & \varphi \land \Next \Always \varphi \\
      \Eventually \varphi & = & \varphi \lor \Next \Eventually \varphi \\
      \varphi \Until \psi & = & \psi \lor (\varphi \land \Next\, (\varphi \Until \psi)) \\
      \varphi \Release \psi & = & \psi \land (\varphi \lor \Next\, (\varphi \Release \psi))
      \end{eqnarray}
      \caption{Important LTL identities}
      \label{fig:identities}
      \end{figure}
\subsection{LTL with Finite Testing}
As can be seen from these examples, LTL makes it easy to specify our application, but actually checking
that our application meets our specification remains a challenge. As \Quickstrom does not have any view of the
application's structure beyond the current trace, we cannot construct a model of the system and apply
the usual LTL model-checking techniques~\citep{ltlmc}. Instead, we randomly explore the state space of
the system by performing randomly-chosen interface actions from a list given in the specification. This gives us
finite, partial traces of the system's execution. LTL, however, is defined on behaviours---infinite, completed traces.
As no finite amount of testing will give an infinitely long trace, we must instead turn to variants of LTL for finite
traces.

The most glaring problem when moving LTL to finite traces is the $\Next$ operator: what does $\Next\varphi$ mean if there
is no next state? Pnueli\footnote{This technique is found in many early papers on LTL with Pnueli as a coauthor such as \citet{fltl}, but \citet{ltlsafety}, which is usually cited for this technique, does not mention finite traces at all.}
answers by splitting the $\Next$ operator into two dual ``next'' operators: The ``weak next'' $\WNext$, which
defaults to $\top$ when there is no next state; and the ``strong next'' $\SNext$, which defaults to $\bot$. The $\Always$ and $\Eventually$ (resp. $\Release$ and $\Until$)
expansion identities then use $\WNext$ and $\SNext$ respectively, so $\Always \varphi$ holds when a violation of $\varphi$ does not occur in the trace, and $\Eventually \varphi$
holds when a state satisfying $\varphi$ occurs at some point in the trace.

Pnueli's finite LTL is still defined for \emph{completed} traces, however. It assumes that the application terminates when the trace ends, and no further states could follow.
By contrast, \Quickstrom traces are partial: they can be extended with more states simply by \Quickstrom further interacting with the application. This means that if we were
to use Pnueli's finite LTL, a liveness property for example about a timer application such as $$\Eventually (\mathsf{timeRemaining} = 0)$$ could be marked as false simply because we didn't wait long enough for the
remaining time to reach zero.

\citet{ltl3} describe a tri-valued LTL for partial traces called LTL$_3$ which distinguishes between those formulae that are evidently true or false only from the trace provided, and those formulae which are \emph{indeterminate}, i.e.\ require
further states to evaluate definitively. \citet{rvltl} later refine LTL$_3$ into RV-LTL, an LTL designed for runtime verification. This logic has four values: formulae may be \emph{definitively} false, such as when a
safety property is shown to be violated; \emph{presumptively false}, such as when a liveness property fails to be fulfilled in the trace; \emph{presumptively true}, such as when
no counterexample to a safety property is found in the trace; or \emph{definitively true}, such as when a liveness property is shown to be satisfied. The definitive cases correspond to the same in LTL$_3$.
In the indeterminate cases, the presumptive results correspond to the answers given by Pnueli's finite LTL.

While RV-LTL is suitable for run-time monitoring or verification, it is still insufficient for testing. Consider our example from earlier
that the menu will not be forever disabled:
$$\Always \Eventually \textsf{menuEnabled}$$
As this formula nests $\Always$ and $\Eventually$ operators, it is definitive in neither positive nor negative cases and will only give presumptive answers.  But the
presumptive answer given in RV-LTL depends only on the value of $\textsf{menuEnabled}$ in the last state of the trace. For a trace where $\textsf{menuEnabled}$ continuously alternates off and on,
the correct presumptive answer would be true, but this formula would be considered presumptively false if we happen to end testing in a state where $\textsf{menuEnabled}$ is false. This would lead to many spurious
counterexamples that, like the liveness property earlier, are merely due to ending our partial trace at the wrong time. Exactly when testing should stop and traces should end to give correct presumptive answers depends
 on the specific formula being tested. Therefore, our \QuickLTL dialect of LTL extends RV-LTL with additional information, allowing users to specify the required length of traces as part of the formula itself.

\subsection{\QuickLTL}

\begin{figure}
  \[
  \begin{array}{lclcll}
\multicolumn{3}{l}{\text{Formulae:}}\\
 \varphi, \psi & ::=  & \multicolumn{3}{l}{ p\ \mid\ \neg \varphi\ \mid\ \top\ \mid\ \bot} \\
    & \mid & \varphi \land \psi &\mid & \varphi \lor \psi \\
  & \mid & \DNext{\varphi} & & & \textit{(required next)} \\
  & \mid & \WNext{\varphi} &\mid& \SNext{\varphi}  & \textit{(weak/strong next)} \\
            & \mid & \Always_n{\varphi} &\mid& \Eventually_n{\varphi}  & \textit{(henceforth/eventually)} \\
            & \mid & \varphi \Until_n \psi  & \mid & \varphi \Release_n \psi  & \textit{(until/release)} \\
            \multicolumn{3}{l}{\text{Guarded form:}}\\
  F,G  &  ::= &  \multicolumn{3}{l}{F \land G\ \mid\  F \lor G}\\
  & \mid &  \multicolumn{3}{l}{\DNext{\varphi}\ \mid\ \WNext{\varphi}\ \mid\ \SNext{\varphi} }\\
  \end{array}
  \]
  \caption{Syntax of \QuickLTL}
  \label{fig:syntax}
  \end{figure}
As can be seen in \Cref{fig:syntax}, we annotate temporal operators with numbers that specify the minimum length of the trace required to give accurate presumptive answers for that operator. For instance, to check $\Always_n \varphi$, \Quickstrom
must check at least $n$ states for $\varphi$ before concluding that the formula is presumptively true; and for $\Eventually_m \psi$ it must check at least $m$ states for $\psi$ before giving up and concluding that the formula is presumptively false.
Adding annotations to our previous example, we get:
$$\Always_{100} \Eventually_{5} \textsf{menuEnabled}$$
These annotations instruct \Quickstrom to check (at least) 100 states for the property $\Eventually_{5} \textsf{menuEnabled}$, which itself requires \Quickstrom to check at least 5 states for $\textsf{menuEnabled}$. These annotations eliminate the
spurious counterexamples mentioned in the previous section, so long as the menu is re-enabled within 5 states of being disabled.
\begin{figure}
  $$
  \begin{array}{lcl}
      \Always_{\ \!0} \varphi & = & \varphi \land \WNext \Always_{\ \!0} \varphi \\
      \Always_{\ \!n+1} \varphi & = & \varphi \land \DNext \Always_{\ \!n} \varphi \\[0.5em]
      \Eventually_{\ \!0} \varphi & = & \varphi \lor \SNext \Eventually_{\ \!0} \varphi \\
      \Eventually_{\ \!n+1} \varphi & = & \varphi \lor \DNext \Eventually_{\ \!n} \varphi \\
  \end{array}
  $$
  \caption{\QuickLTL expansions for basic temporal operators.}
  \label{fig:quickltl_expansions}
\end{figure}
The semantics of these annotations is best explained by their expansions into the ``next'' operators, given in \Cref{fig:quickltl_expansions}. In addition to the ``weak next'' $\WNext$ and ``strong next'' $\SNext$ of RV-LTL, we also introduce the self-dual ``required next'' $\DNext$, which, rather
than default to a value in the absence of a next state, simply requires \Quickstrom to perform more actions to \emph{produce} a next state if one does not exist. As can be seen in \Cref{fig:quickltl_expansions}, the numeric annotation $n$ on a temporal operator expands into $n$ uses of the $\DNext$ operator,
thus requiring \Quickstrom to generate and check at least $n$ states to evaluate the formula for that operator.

\subsection{Evaluation by Formula Progression}

We evaluate \QuickLTL formulae in a step-by-step manner, unrolling and partially evaluating the formula
for each state of the trace in succession, similar to an operational semantics but for LTL formulae.
Evaluation of a formula $\varphi$ proceeds in three phases, repeated in a loop:
\begin{enumerate}
  \item Given the state $\sigma$, unroll the formula $\varphi$ one step and partially evaluate it against $\sigma$, according to the rules given in \Cref{fig:evaluation}. This relation $\varphi \xmapsto{\sigma} \varphi'$ evaluates all atomic propositions about the state $\sigma$, leaving a formula $\varphi'$ where
  all nontrivial propositions are surrounded by a ``next'' operator. Note that the rules for temporal operators are expanding formulae exactly as in the expansion identities of \Cref{fig:quickltl_expansions}.

  \begin{figure}
    \begin{gather*}
      \boxed{\varphi \xmapsto{\sigma} \psi}\\
      \inferrule{ }{\top \xmapsto{\sigma} \top}\quad\inferrule{ }{\bot \xmapsto{\sigma} \bot}\\
      \inferrule{ }{p \xmapsto{\sigma} p(\sigma)}\quad
      \inferrule{\varphi \xmapsto{\sigma} \varphi' }{\neg \varphi \xmapsto{\sigma} \neg \varphi'}
      \\
      \inferrule{\varphi \xmapsto{\sigma} \varphi' \\ \psi \xmapsto{\sigma} \psi'}{\varphi \land \psi \xmapsto{\sigma} \varphi' \land \psi'}\quad
      \inferrule{\varphi \xmapsto{\sigma} \varphi' \\ \psi \xmapsto{\sigma} \psi'}{\varphi \lor \psi \xmapsto{\sigma} \varphi' \lor \psi'}\\
      \inferrule{ }{\DNext \varphi \xmapsto{\sigma} \DNext \varphi}\quad
      \inferrule{ }{\SNext \varphi \xmapsto{\sigma} \SNext \varphi}\quad
      \inferrule{ }{\WNext \varphi \xmapsto{\sigma} \WNext \varphi}\\
      \inferrule{ \varphi \xmapsto{\sigma} \varphi' }{ \Always_{n+1} \varphi \xmapsto{\sigma} \varphi' \land \DNext \Always_{n} \varphi }\quad
      \inferrule{ \varphi \xmapsto{\sigma} \varphi' }{ \Always_{0} \varphi \xmapsto{\sigma} \varphi' \land \WNext \Always_{0} \varphi }\\
      \inferrule{ \varphi \xmapsto{\sigma} \varphi' }{ \Eventually_{n+1} \varphi \xmapsto{\sigma} \varphi' \lor \DNext \Eventually_{n} \varphi }\quad
      \inferrule{ \varphi \xmapsto{\sigma} \varphi' }{ \Eventually_{0} \varphi \xmapsto{\sigma} \varphi' \lor \SNext \Eventually_{0} \varphi }\\
      \inferrule{ \varphi \xmapsto{\sigma} \varphi' \\ \psi \xmapsto{\sigma} \psi' }{ \varphi \Until_{n+1} \psi \xmapsto{\sigma} \psi' \lor (\varphi' \land \DNext\ \!(\varphi \Until_{n} \psi)) }\\
      \inferrule{ \varphi \xmapsto{\sigma} \varphi' \\ \psi \xmapsto{\sigma} \psi' }{ \varphi \Until_{0} \psi \xmapsto{\sigma} \psi' \lor (\varphi' \land \SNext\ \!(\varphi \Until_0 \psi)) }\\
      \inferrule{ \varphi \xmapsto{\sigma} \varphi' \\ \psi \xmapsto{\sigma} \psi' }{ \varphi \Release_{n+1} \psi \xmapsto{\sigma} \psi' \land (\varphi' \lor \DNext\ \!(\varphi \Release_{n} \psi)) }\\
      \inferrule{ \varphi \xmapsto{\sigma} \varphi' \\ \psi \xmapsto{\sigma} \psi' }{ \varphi \Release_{0} \psi \xmapsto{\sigma} \psi' \land (\varphi' \lor \WNext\ \!(\varphi \Release_{0} \psi)) }
    \end{gather*}
    \caption{Unrolling a formula, evaluating it against one state}
    \label{fig:evaluation}
  \end{figure}
  \item  Simplify the resultant formula $\varphi'$ using simple logical identities and the negation identities 1--5 from \Cref{fig:identities}. This will either result in a definitive answer like $\top$ or $\bot$, in which case \Quickstrom will cease checking; or it will result in a formula $F$ in \emph{guarded form}, syntax of
  which is given in \Cref{fig:syntax}. A formula is in guarded form if it consists solely of conjunctions and disjunctions of formulae guarded by ``next'' operators. If none of these ``next'' operators are the ``required next'' $\DNext$, then a presumptive answer can be given by treating all $\WNext$-guarded terms as $\top$ and all $\SNext$-guarded terms as $\bot$, then simplifying the formula.

  \item If the guarded-form formula $F$ contains $\DNext$-guarded terms, then \Quickstrom must perform more actions to generate a new state $\sigma$. We then step the formula forward according to the rules in \Cref{fig:stepping}. This relation $F \Mapsto \varphi$ progresses the formula
  to the next state, giving a new formula $\varphi$ that can be used in the next iteration of the loop with the new state $\sigma$.

\begin{figure}
  \begin{gather*}
    \boxed{G \Mapsto \varphi}\\
    \inferrule{G \Mapsto \varphi \\ F \Mapsto \psi}{G \land F \Mapsto \varphi \land \psi }\quad
    \inferrule{G \Mapsto \varphi \\ F \Mapsto \psi}{G \lor F \Mapsto \varphi \lor \psi }\\
    \inferrule{ }{\DNext \varphi \Mapsto \varphi }\quad \inferrule{ }{\SNext \varphi \Mapsto \varphi }\quad\inferrule{ }{\WNext \varphi \Mapsto \varphi }
  \end{gather*}
  \caption{Stepping a formula forward}
  \label{fig:stepping}
  \end{figure}
\end{enumerate}
This procedure is very similar to the \emph{formula progression} technique proposed by Kabanza et al.~\citep{progress1, progress2} for standard LTL,
however we split the progression relation into the two relations in \Cref{fig:evaluation,fig:stepping} to allow us to distinguish between formulae that already have definitive answers and those that have only presumptive answers
due to the presence of remaining ``next'' operators.

\citet{rosu} warn that this technique can result in
exponential blow-up in the size of the formula relative to the number of steps taken,  however we have found in our case studies that this is avoided in all
practical cases by our simplification of the formula at each step. Nested temporal operators can cause the formula size to grow at each step but, as our traces are rarely
longer than a few hundred states, this is not prohibitively expensive. Therefore, this technique remains effective for our practical scenarios.

\section{\Specstrom}\label{sec:specstrom}

\QuickLTL formulae are only a small component of a \Quickstrom specification. Specification writers must also describe which \emph{actions} can occur, either due to \Quickstrom interacting directly with the interface or
due to asynchronous environmental events, as well as the \emph{state queries} that make up the atomic propositions in a \QuickLTL formula.

Our specification language, \Specstrom, is a simple language with syntax that superficially resembles JavaScript and has smooth interoperability with JavaScript data such as objects and arrays, but with a significantly more
restricted semantics: recursion is not allowed, and all expressions terminate.
\Specstrom also includes a number of built-in primitives for constructing formulae, actions and state queries.

Although \Specstrom supports higher order functions, it still guarantees termination through use of a very simple type system.
Because most web programmers are not accustomed to strict type-checking, this type system is designed to be mostly invisible to
the programmer: it distinguishes only between \emph{functions} and \emph{non-functions}, and all types are inferred. To avoid circumvention
of this type system, functions may not be
placed inside data types such as arrays or objects. The termination guarantee that we obtain from this type system enables us to
analyse \Specstrom code more easily, as in \Cref{ssec:staticanalysis}.
\subsection{Evaluation Control}
\Specstrom also gives the user fine-grained control over evaluation, allowing programmers to define their own temporal operators or connectives.
As an illustrative example, consider the following temporal predicate $\mathsf{evovae}(x)$, which states that $x$ shall forever have the same value it had initially:
$$
\mathsf{evovae}(x) = \textbf{let}\ v = x; \Always(x\ \texttt{==}\ v);
$$
In a language which evaluates in applicative order (i.e. "strict" evaluation), this would trivially be true, because the parameter $x$ would be fully evaluated to a value before $\mathsf{evovae}$ was even invoked, and thus
$x\ \texttt{==}\ v$ would be true independently of the state in which it is executed. On the other hand, in a language where bindings are only evaluated when they are used, this would \emph{also} be trivially true, because binding
$v$ to $x$ would not evaluate $x$ until inside the $\Always$ operator, making $\mathsf{evovae}(x)$ equivalent to the trivial $\Always (x\ \texttt{==}\ x)$. For this reason, \Specstrom allows the user
to specify which bindings are to be left unevaluated explicitly, with a $\sim$ prefix before the binding. This allows us to define $\mathsf{evovae}$ with the intended semantics, where only $x$ is left
unevaluated and $v$ is evaluated eagerly:
$$
\mathsf{evovae}(\sim \!x) = \textbf{let}\ v = x; \Always(x\ \texttt{==}\ v);
$$
\subsection{Specifying an Egg Timer}
\begin{figure}
  $$
  \begin{array}{l}
  \textbf{let} \sim\!\!\mathit{stopped} = {}^\backprime\texttt{\#toggle}^\backprime.\mathsf{text}\ \texttt{==}\ \texttt{"start"};\\
  \textbf{let} \sim\!\!\mathit{started} = {}^\backprime\texttt{\#toggle}^\backprime.\mathsf{text}\ \texttt{==}\ \texttt{"stop"};\\
  \textbf{let} \sim\!\!\mathit{time} = \mathsf{parseInt}({}^\backprime\texttt{\#remaining}^\backprime.\mathsf{text});\\[0.5em]
  \textbf{action}\ \mathsf{start!} = \mathsf{click}!(^\backprime\texttt{\#toggle}^\backprime)\ \textbf{when}\ \mathit{stopped};\\
  \textbf{action}\ \mathsf{stop!} = \mathsf{click}!(^\backprime\texttt{\#toggle}^\backprime)\ \textbf{when}\ \mathit{started};\\
  \textbf{action}\ \mathsf{wait!} = \mathsf{noop}!\ \textbf{timeout}\ 1000\ \textbf{when}\ \mathit{started};\\
  \textbf{action}\ \mathsf{tick?} = \mathsf{changed}?({}^\backprime\texttt{\#remaining}^\backprime);\\[0.5em]
  \textbf{let} \sim\!\!\mathit{ticking}\ \{ \\ \quad \textbf{let}\ \mathit{old} = \mathit{time};
  \\ \quad  \mathit{started}
  \\ \qquad \land\ \WNext\ ( \mathsf{tick?}\ \mathsf{in}\ \mathrm{happened}\\\ \ \qquad\quad \land\ \mathit{time}\ \texttt{==}\ \mathit{old} - 1 \\\  \ \qquad\quad \land\ \textbf{if}\ \mathit{time}\ \texttt{==}\ 0\ \{ \mathit{stopped} \}\ \textbf{else}\ \{ \mathit{started} \}) \\ \}\\
  \textbf{let} \sim\!\!\mathit{waiting} =  \\\quad \mathit{started} \land \WNext (\mathsf{wait!}\ \mathsf{in}\ \mathrm{happened} \land \mathit{started});\\
  \textbf{let} \sim\!\!\mathit{starting} = \\
  \quad \mathit{stopped} \land \WNext\ (\mathsf{start!}\ \mathsf{in}\ \mathrm{happened}\\
  \qquad\qquad\quad \quad\!\! \land\ \textbf{if}\ \mathit{time}\ \texttt{==}\ 0\ \{ \mathit{stopped} \}\ \textbf{else}\ \{ \mathit{started} \});\\
  \textbf{let} \sim\!\!\mathit{stopping} = \\\quad \mathit{started} \land \WNext (\mathsf{stop!}\ \mathsf{in}\ \mathrm{happened} \land \mathit{stopped});\\[0.5em]
  \textbf{let} \sim\!\!\mathit{safety} = \\
  \quad \mathsf{loaded?}\ \mathsf{in}\ \mathrm{happened} \land  \mathit{time}\ \texttt{==}\ 180
  \\\qquad \land\ \Always_{400} (\mathit{starting} \lor \mathit{stopping} \lor \mathit{waiting} \lor \mathit{ticking});\\
  \textbf{let} \sim\!\!\mathit{liveness} = \\\quad \Always_{400} (\mathsf{start!}\ \mathsf{in}\ \mathrm{happened} \Rightarrow \Eventually_{360} \mathit{stopped}); \\
  \textbf{let} \sim\!\!\mathit{timeUp} = \\\quad \Always_{400} (\mathsf{start!}\ \mathsf{in}\ \mathrm{happened} \Rightarrow \Eventually_{360}(\mathit{time}\ \mathtt{==}\ 0)); \\[0.5em]
  \textbf{check}\ \mathit{safety}\ \mathit{liveness};\\
  \textbf{check}\ \mathit{timeUp}\ \textbf{with}\ \mathsf{start!}\ \mathsf{wait!}\ \mathsf{tick?};
  \end{array}
  $$
  \caption{An Example of a \Specstrom specification}
  \label{fig:timer}
  \end{figure}

\Cref{fig:timer} gives an illustrative example of a complete \Specstrom specification for a three-minute egg timer application, with syntax slightly adjusted for brevity. The application consists of a start/stop toggle button and a label
containing the remaining time in seconds.

\paragraph{State projections} The first two lines introduce atomic propositions, $\mathit{stopped}$ and $\mathit{started}$, which indicate the status of the timer.
 Strings surrounded in ${}^\backprime\texttt{backticks}^\backprime$ are \emph{CSS selectors} which extract part of the application's UI state. In this case, our propositions
 are determined by the text label on the toggle button. Note that these definitions are expected to change over time and are thus bound with the $\sim$ operator to prevent them
 from being evaluated at definition-time. We similarly define a state-dependent quantity $\mathit{time}$ which is determined from the label containing the number of seconds remaining.

 \paragraph{Actions} The next four lines define \emph{actions} that may occur, either due to \emph{user actions}, which are initiated by \Quickstrom, or due to \emph{events}, which are asynchronously initiated
 by the application. In our specifications and libraries, we adopt the convention that user actions are suffixed with an exclamation mark (!) and events with a question mark (?).
 Thus, we define three user actions and one event. The event is $\mathsf{tick?}$, which fires when the application updates the remaining time label each second.
 The user actions are $\mathsf{start!}$, $\mathsf{stop!}$ and $\mathsf{wait!}$, which all indicate actions \Quickstrom may take when
 interacting with the application. They are defined in terms of built-in primitive actions such as $\mathsf{noop!}$ and $\mathsf{click!}()$. We associate \emph{guards} to our actions using
 the $\textbf{when}$ operator, allowing us to differentiate between the $\mathsf{start!}$ and $\mathsf{stop!}$ actions: both of these actions simply involve clicking the toggle button, but in different contexts.
 In general, these guards take the form of atomic propositions, i.e.\ non-temporal boolean formulae. The action will only fire if the guard condition is met.

 \paragraph{Timeouts} The definition for the action $\mathsf{wait!}$ associates a \emph{timeout} to the built-in action $\mathsf{noop!}$ with the $\mathbf{timeout}$ keyword.
 This indicates to \Quickstrom that, after performing this action, it should not attempt to perform another action for at least one second, or until an event occurs. These timeouts
 are designed to accommodate the very common use case where a user action causes an application to respond asynchronously. In this case, the action $\mathsf{noop!}$ does nothing, so
 the action $\mathsf{wait!}$ will cause \Quickstrom to wait until a $\mathsf{tick?}$ occurs, or until one second elapses. This action is needed because otherwise \Quickstrom will simply stop the timer as soon as it starts, as
 it has no other actions available to perform once the timer has started.

 \paragraph{Safety properties} The next five definitions can be understood by looking at the property \emph{safety}.
This property states that the built-in event $\mathsf{loaded?}$ must happen first, and that in the initial state,
 the time remaining should be three minutes. It then states that one of $\mathit{ticking}, \mathit{waiting}, \mathit{starting}$, and $\mathit{stopping}$ must
 be true forevermore. Each of these properties describes one allowable state transition. For example, $\mathit{stopping}$ describes a transition from a state where the timer
 was $\mathit{started}$ to a state where the timer has $\mathit{stopped}$ due to the $\mathsf{stop!}$ action. The variable ``happened'' is a special state-dependent variable that contains
 all events or actions that occurred immediately prior to the current state. The $\mathit{ticking}$ transition uses a $\textbf{let}$-binding to
 freeze the value of $\mathit{time}$ before the $\mathsf{tick?}$ event occurred. This allows us to then specify that the value of $\mathit{time}$ after the event must be decremented.
 \begin{figure*}
  \begin{tabular}{ll}
  $\textbf{Checker}$&$\textbf{Executor}$\\
  $\quad\mathsf{Start}\ \langle\mathit{dependencies}\rangle$ & $\quad\mathsf{Event}\ \langle\mathit{event}\rangle\ \langle\mathit{state}\rangle$ \\
  $\qquad$ \emph{Request a new session be started} & $\qquad$ \emph{Notify the checker about an event that occurred}\\
  $\qquad$ \emph{(also specifies which selectors are relevant)} & $\qquad$ \emph{along with the updated state}\\[0.5em]
  $\quad\mathsf{Act}\ \langle\mathit{action}\rangle\ \langle\mathit{version}\rangle\ \langle\mathit{timeout}\rangle$ & $\quad\mathsf{Acted}\ \langle\mathit{state}\rangle$\\
  $\qquad$ \emph{Request the given action be performed} & $\qquad$ \emph{Notify the checker that an action was performed} \\
  $\qquad$ \emph{(rejected if version $<$ trace length)} & $\qquad$ \emph{along with the updated state}\\[0.5em]
  $\quad\mathsf{Wait}\ \langle\mathit{time}\rangle\ \langle\mathit{version}\rangle$ & $\quad\mathsf{Timeout}\ \langle\mathit{state}\rangle$ \\
  $\qquad$ \emph{Request to signal a \textsf{Timeout} after the given time} & $\qquad$ \emph{Notify the checker that a timeout has elapsed} \\
  $\qquad$ \emph{if no event occurs first.} & $\qquad$ \emph{along with the (possibly) updated state}
  \end{tabular}
  \caption{The protocol between the checker and executor.}
\label{fig:comms2}
\end{figure*}
 \paragraph{Liveness properties} While the safety property defined above thoroughly describes what transitions are allowable, it does not say anything about what
 transitions will be taken. For this, we need liveness properties. The simplest liveness property of our egg timer is that the timer will eventually stop---either by running out of time or by the user pressing the stop button.
 This is easily expressed in the property $\mathit{liveness}$. For an egg timer, however, it is reasonable to want a stronger property: eventually, time will run out. Unfortunately this property, which we call $\mathit{timeUp}$ in \Cref{fig:timer}, is not necessarily
 true for all implementations. For example, if the timer is implemented with a one-second granularity, the user might repeatedly press the start and stop button faster than the granularity of the timer, and prevent it
 from ever making progress. An optional parameter to the $\mathbf{check}$ command, seen at the bottom of \Cref{fig:timer}, allows us to specify which actions may fire when testing a given property.
 Therefore, we can still check this property by excluding the $\mathsf{stop!}$ action from the set of allowable actions. Then, the only way the timer will stop is if it runs out of time.

 Currently, the \Quickstrom checker makes a completely random selection from the set of allowable actions for the current state. Refining this action selection to be more \emph{targeted}, methodically exploring previously unreached 
 parts of the state space, is left as future work (see \Cref{ssec:targeted}). 

 \subsection{Static Analysis}\label{ssec:staticanalysis}
\Quickstrom is built on top of Selenium WebDriver~\citep{selenium}, a programmatic testing tool which can simulate user interaction with a web application using a headless browser instance.
When given a \Specstrom specification, \Quickstrom must determine what parts of the browser state are relevant for the properties at hand \emph{before} checking, to properly instrument the running application with listeners for
changes to relevant components of the user interface, and to get a consistent view of a state by retrieving all relevant information in bulk. We determine this information automatically by statically
analysing \Specstrom code. Because \Specstrom guarantees termination and does not support recursion, this analysis is a very simple abstract interpretation for dependency analysis. In addition to direct dependencies,
such as the expression ${}^\backprime\texttt{\#toggle}^\backprime.\mathsf{text}$ which depends obviously on the UI element ${}^\backprime\texttt{\#toggle}^\backprime$, we must also track indirect dependencies, such as in the
expression
$\textbf{if}\ \; \!{}^\backprime\texttt{\#toggle}^\backprime.\mathsf{enabled}\ \{ 0 \}\ \textbf{else}\ \{ 1 \}$, which also depends on ${}^\backprime\texttt{\#toggle}^\backprime$. Running this analysis on the property under test
yields a set of state elements which are instrumented and recorded by \Quickstrom as it runs actions.

\subsection{Checker and Executor}

\begin{figure}
  \begin{tikzpicture}
    \draw[fill=green!5!white,rounded corners=0.1em] (0,-2) rectangle (2,5);
    \draw[fill=red!5!white, rounded corners=0.1em] (6,-2) rectangle (8,5);
    \node at (7,1.5) {\textbf{Executor}};
    \node at (1,1.5) {\textbf{Checker}};
    \draw[->,draw=blue!60!black,  thick] (2,4.75)--(6,4.5) node[midway,above]{$\mathsf{Act}\ \textsf{click!}\ 0$};
    \draw[<-,draw=green!60!black,  thick] (2,3.5)--(6,3.75) node[midway,above]{$\mathsf{Acted}\ \langle\textit{state}\rangle$};
    \draw[<-,draw=blue!60!black,  thick] (2,2.5)--(6,2.75) node[midway,above]{$\mathsf{Event}\ \textsf{changed?}\ \langle\textit{state}\rangle$};
    \draw[->,draw=blue!60!black,  thick] (2,1.75)--(6,1.5) node[midway,above]{$\mathsf{Act}\ \textsf{pressKey!}\ 2$};
    \draw[<-,draw=green!60!black,  thick] (2,0.5)--(6,0.75) node[midway,above]{$\mathsf{Acted}\ \langle\textit{state}\rangle$};
    \draw[<-,draw=blue!60!black,  thick] (3,-0.75)--(6,-0.5) node[midway,xshift=-1.5em,above]{$\mathsf{Event}\ \textsf{changed?}\ \langle\textit{state}\rangle$};
    \draw[->,draw=red!60!black,  thick] (2,-1.5)--(6,-1.75) node[midway,above]{$\mathsf{Act}\ \textsf{pressKey!}\ 3$};
\end{tikzpicture}
\caption{Example of communication between checker and executor.}
\label{fig:comms}
\end{figure}
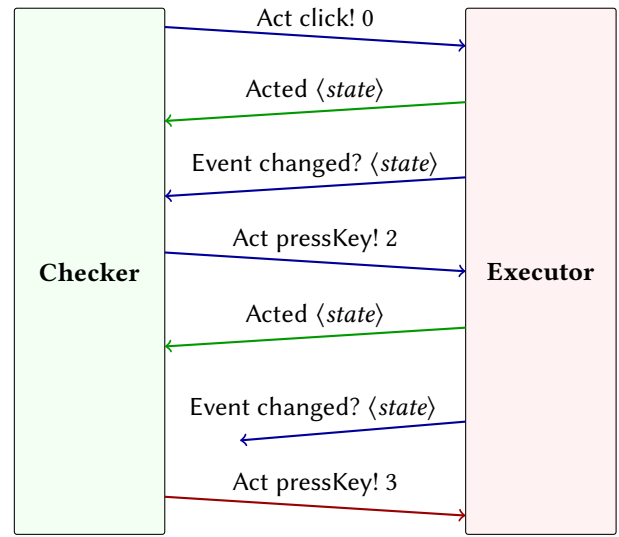

\Quickstrom is divided into two main components: the \emph{checker}, which is the \Specstrom interpreter that evaluates the formula and selects actions to perform; and the \emph{executor}, which interacts with Selenium WebDriver to
actually interact with the application under test using a headless browser instance.

\Cref{fig:comms2} describes the protocol for communication between the checker and executor. Each column describes the messages sent by the checker and executor respectively.
 When the checker intends to test a property, it signals the executor to load the application (\textsf{Start}) and tells it which parts of the application's state are relevant to the
property. As mentioned, this information is determined by simple static analysis of the \Specstrom code.  The executor uses this information to add event listeners to the application under test, and to determine what parts of the state to include in future messages.
The executor then waits for events to occur in the application or action requests (\textsf{Act}) to come from the checker. In either case it reports the updated application state to the checker, after performing the requested action if necessary (\textsf{Acted} and \textsf{Event}).  The checker will wait until the initial event of the property is observed (usually, this is when the page is \textsf{loaded}) before
beginning to request actions.

For user actions that include a $\textsf{timeout}$, such as our $\mathsf{wait!}$ action from the egg timer example, the \textsf{Act} message
may optionally include a $\textbf{timeout}$ parameter. If the specified time elapses without an event occurring, the executor will send a \textsf{Timeout} message
along with the state (which might have changed since the last action occurred). The checker can also request such a timeout separately from an action
using the \textsf{Wait} message, which is used when a $\textbf{timeout}$ is associated with an event: if the event occurs, the checker requests a timeout from the executor.

Because the application under test is running in a separate process and cannot be paused, it is possible that asynchronous events could change the application's state while the checker is deciding what action to perform. Thus, the checker
might make a decision based on out-of-date information. We solve this problem by including the \emph{length of the trace so far} in every message after checking begins. \Cref{fig:comms} illustrates an interaction between the checker and executor after initiating a run for a particular property. Time flows
from the top to the bottom.
Initially, the checker tells the executor to \textsf{click} a button, which the executor dutifully does, returning the updated state
along with its acknowledgment that the action was performed. Then, part of the application's state is asynchronously \textsf{changed},
which the executor reports to the checker along with an updated state. The checker acknowledges receipt of all these updated states by
including the current trace length ($2$) in its next action request, to press a key. The executor then performs this action and sends the new state to
the checker. Then, the application state is again asynchronously \textsf{changed}, but before the checker is notified of this, it requests that the executor
perform another action to press a key. Because this request has an out of date trace length ($3$, not $4$), the executor knows to ignore this request.

Nothing about the \emph{checker} is specific to Selenium WebDriver: paired with a different executor, the same checker could be used to test any reactive system.
While the only production-ready executor is the WebDriver-based one, to simplify testing of our \Specstrom interpreter we have also implemented another executor, which interprets models written in Milner's Calculus of Communicating Systems~\citep{ccs}.
Developing other executors is promising future work.

\section{Evaluation}\label{sec:todomvc}
\begin{figure}
  \centering
\frame{\includegraphics[width=0.4\textwidth]{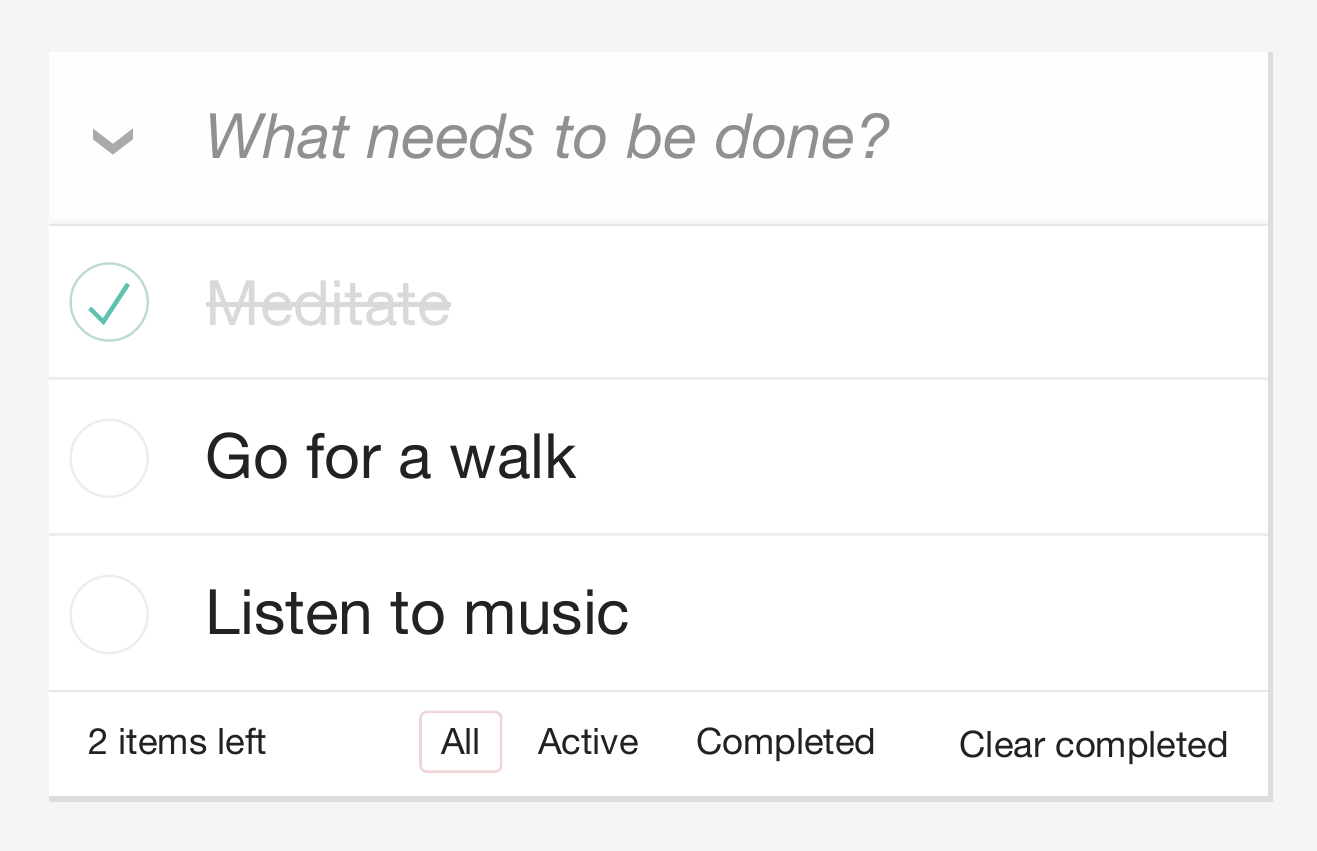}}
\caption{A TodoMVC implementation in action.}
\label{fig:sshot}
\end{figure}
The TodoMVC benchmark is a suite of various implementations of the same to-do list application, which should all look the same and behave according
to the same (plain English) specification~\citep{todomvc}. The various implementations are provided by independent developers, usually the developers of the frameworks themselves.
The purpose of the benchmark is to provide a non-trivial application that can be used to compare frameworks for performance, functionality and ease of use.

\Cref{fig:sshot} contains a screenshot of one of the TodoMVC implementations in action. As can be seen, items can be added by typing into the text box at the top of the list, and items may be
marked ``completed'' by clicking the checkbox to their left. The arrow icon to the left of the text entry box allows all items to be toggled simultaneously. Items may be filtered by their status
using the buttons below the list, and items may be edited by double clicking on them. A delete button appears to the right of an item when the user hovers over it.
The to-do list is \emph{persistent}, stored in local storage, so page reloads should not affect the content of the to-do list.

\subsection{A Formal TodoMVC Specification}
The TodoMVC specification is quite precise, but it is written in technical English, not a formal specification language.
Therefore, we have translated the TodoMVC specification into a formal specification, consisting of 300 lines of \Specstrom.
As has been observed with other natural language specifications~\citep{lsrp}, our formalisation efforts show the official specification is rife with ambiguity and under-specification.

    For instance, the official TodoMVC specification defines what items should be shown when the
    user changes the current filter, but it does not say what happens to the \emph{rest} of the user interface.
    Our formal specification makes the reasonable assumption that no other part of the interface (such as pending input) should be modified when switching between
    filters, even though the official specification does not explicitly rule out such behaviour.

    The official specification also says nothing about which filter should be
    active after all to-do items have been removed, and
    our specification does the same, i.e.\ it leaves it undefined. Interestingly, there
    seems to be a commonly understood de-facto specification that the filter should be unchanged---developers
    have submitted bug-fix pull requests to implementations that behave differently---but we have not
    formalised this as it is not officially specified by TodoMVC.
\begin{figure}
  $$
\begin{array}{l}
  \textbf{let} \sim\!\!\mathit{safety} = \mathit{initial} \\
  \quad\!\!\!\begin{array}{rl} \land\ \Always\ (\ \ & \!\!\!\!\mathit{focusNewTodo} \\
    \lor & \!\!\!\!\mathit{enterNewTodoText} \\
    \lor & \!\!\!\!\mathit{addNew} \\
    \lor & \!\!\!\!\mathit{changeFilter} \\
    \lor & \!\!\!\!\mathit{setSameFilter} \\
    \lor & \!\!\!\!\mathit{toggleAll} \\
    \lor & \!\!\!\!\mathit{checkOne} \\
    \lor & \!\!\!\!\mathit{uncheckOne} \\
    \lor & \!\!\!\!\mathit{delete} \\
    \lor & \!\!\!\!\mathit{enterEditMode} \\
    \lor & \!\!\!\!\mathit{inEditMode} \;
    )\ %
  \end{array}\\
  \quad\land\ \Always \cdots \langle \text{invariants} \rangle \\
  \textbf{let} \sim\!\!\mathit{enterEditMode} = \mathit{startEditing} \land \WNext \mathit{editMachine} \\
  \textbf{let} \sim\!\!\mathit{editMachine}\ \{ \\
    \quad \textbf{let}\ \mathit{item} = \mathit{itemInEditMode};\\
  \quad \mathit{exitEditMode}(\mathit{item})\ \Release\\\qquad ( \mathit{enterEditText} \lor \mathit{exitEditMode}(\mathit{item}) )\\
  \}\\
  \textbf{let}\ \mathit{exitEditMode}(\mathit{initialItem}) =\\
  \qquad\;\;\! \!  \mathit{commitEdit}(\mathit{initialItem})\\
  \quad\  \lor\ \mathit{abortEdit}(\mathit{initialItem})
\end{array}
$$
  \caption{Sketch of our TodoMVC specification}
  \label{fig:todomvcspec}
\end{figure}

\Cref{fig:todomvcspec} gives a high-level sketch illustrating the main safety property for TodoMVC in our formal specification. When numeric subscripts on temporal operators
are omitted, they use a user-specified default value. Using a higher value increases test accuracy but also test running time. See \Cref{ssec:timing} for a detailed analysis 
of this trade-off.
Like our timer specification, we specify the application similarly to a state machine. The property consists of three conjuncts: one specifying the initial state,
one specifying the allowable transitions corresponding to user actions, and one containing a list of invariants. These invariants mostly just state the requirement
that the various elements that make up the user interface are actually present.

This kind of state machine specification is a very common pattern when writing \Quickstrom specifications, and our TodoMVC example also demonstrates another pattern: We can use the temporal
operator $\Release$ to \emph{nest} these state machine specifications. Notice that the transition conjunct of the main safety property is easily satisfied if we are editing an item (i.e.\ the $\mathit{inEditMode}$ disjunct is true).
This is because we specify editing an item as a separate state machine specification in $\mathit{editMachine}$, which is invoked by the $\mathit{enterEditMode}$ transition.  From the perspective of the main, high-level state machine in $\mathit{safety}$, editing an item
is a single abstract state with a single internal transition, described by the formula $\mathit{inEditMode}$. But in $\mathit{editMachine}$, we refine this into three transitions: editing the text of an item, committing changes, and aborting changes.
If either of those last two transitions are taken, we leave the nested state machine: we are no longer in edit mode and need no longer abide by the nested
state machine specification.  We use the release operator $\Release$ to indicate this (Note that the top-level $\mathit{safety}$ state machine uses $\Always$, which is equivalent to $\Release$ with an exit condition of $\bot$).
In addition, we also ``remember'' the original value of an item that is being edited, using the $\textbf{let}$ binding in $\mathit{editMachine}$, so that we can specify that the text returns to its original value if an edit is aborted.

    We have not yet formalized the persistence aspect of the official specification. We expect that this could be modelled by
    inserting page reloads as another possible action, and may expose further problems in the implementations' handling of local storage,
    but this is left as future work.

\begin{table}
  \begin{tabular}{|l|lll}\hline
    \textbf{Passed} --- \textbf{23} (\textit{9 beta}, 14 mature)
    \\\hline
    \textit{angularjs\textunderscore{}require},
    \textit{aurelia},
    \textit{backbone\textunderscore{}require},
    backbone,\\
    binding-scala,
    closure,
    emberjs,
    \textit{enyo\textunderscore{}backbone},\\
    \textit{exoskeleton},
    js\textunderscore{}of\textunderscore{}ocaml,
    \textit{jsblocks},
    knockback,\\
    knockoutjs,
    kotlin-react,
    \textit{react-alt},
    \textit{react-backbone},\\
    react,
    \textit{riotjs},
    scalajs-react,
    typescript-angular,\\
    typescript-backbone,
    typescript-react,
    vue
    \\\hline
    \textbf{Failed} --- \textbf{20} (\textit{8 beta}, 12 mature)
    \\\hline
    
    angular-dart\hyperlink{p11}{\textsuperscript{14}},
    \textit{angular2\textunderscore{}es2015\hyperlink{p5}{\textsuperscript{1}}},
    \textit{angular2\hyperlink{p3}{\textsuperscript{5}}},
    angularjs\hyperlink{p4}{\textsuperscript{7}},\\
    backbone\textunderscore{}marionette\hyperlink{p9}{\textsuperscript{11}},
    \textit{canjs\textunderscore{}require}\hyperlink{p14}{\textsuperscript{13}},
    canjs\hyperlink{p14}{\textsuperscript{13}},
    \textit{dijon\hyperlink{p6}{\textsuperscript{2}}},\\
    dojo\hyperlink{p2}{\textsuperscript{9}},
    \textit{duel\hyperlink{p4}{\textsuperscript{4}}},
    elm\hyperlink{p1}{\textsuperscript{4}},
    jquery\hyperlink{p12}{\textsuperscript{10}},
    \textit{knockoutjs\textunderscore{}require\hyperlink{p6}{\textsuperscript{2}}},\\
    \textit{lavaca\textunderscore{}require\hyperlink{p4}{\textsuperscript{4}}},
    mithril\hyperlink{p4}{\textsuperscript{7}},
    polymer\hyperlink{p10}{\textsuperscript{6}},
    \textit{ractive}\hyperlink{p13}{\textsuperscript{12}}, reagent\hyperlink{p1}{\textsuperscript{4}},\\
    vanilla-es6\hyperlink{p7}{\textsuperscript{8,}}\hyperlink{p8}{\textsuperscript{3}},
    vanillajs\hyperlink{p7}{\textsuperscript{8}}\\
    \hline
  \end{tabular}
  \caption{Summary of Results}
  \label{tbl:Results}
\end{table}

\subsection{Results}
From the many standard TodoMVC 1.3 implementations listed on the TodoMVC website~\citep{todomvc}, we selected 43 implementations for our evaluation.
We selected only those implementations that are stored on the TodoMVC repository (commit version \texttt{41ba86d} from February 2020)
to ensure reproducibility. We also excluded any implementations that were not standard, single-page TodoMVC applications (e.g. streaming variants such as those based on Firebase), those
that didn't successfully start (i.e.\ \texttt{cujo}), those whose markup didn't match the specification (i.e.\ \texttt{gwt}), and those for whom compiled, testable artifacts were not available (i.e.\ \texttt{react-hooks}, \texttt{emberjs-require}).
   Some of these implementations are labelled as \emph{beta}, i.e.\ still under evaluation from the TodoMVC team.
As can be seen in \Cref{tbl:Results}, which gives a high level overview of our results on each of these implementations, we found bugs or faults in 20 of those implementations---almost half. Surprisingly, this fault rate was not significantly higher
for the implementations marked as beta, although bugs due to missing features are more common.

\begin{table}
  \begin{tabular}{|r|p{6cm}|r|}\hline
    & \textbf{Description} & \textbf{Count} \\\hline
    \hypertarget{p5}{1}&  Items have no checkboxes& 1\\\hline
    \hypertarget{p6}{2}& There are no filter controls& 2 \\\hline
    \hypertarget{p8}{3}& A \texttt{<strong>} element is missing& 1\\\hline
    \hypertarget{p1}{4}& Blank items can be added & 1 \\\hline
    \hypertarget{p3}{5}&  Edit input is not focused after double-click& 1 \\\hline
    \hypertarget{p10}{6}& Incorrectly pluralizes the to-do count text& 1\\\hline
    \hypertarget{p4}{7}& Any pending input is cleared on filter change or removal of last item& 4\\\hline
    \hypertarget{p7}{8}& A new item is created from pending input after non-create actions& 2\\\hline
    \hypertarget{p2}{9}& ``Toggle all'' does not untoggle all items when certain filters are enabled& 1\\\hline
    \hypertarget{p12}{10}& The ``Toggle all'' button disappears when the current filter contains no items. & 1\\\hline
    \hypertarget{p9}{11}& Commiting an empty to-do item in edit mode does not fully delete it---it can later be restored with ``Toggle all''& 1\\\hline
    \hypertarget{p13}{12}& Editing an item hides other items& 1\\\hline
    \hypertarget{p14}{13}& Adding an item changes the filter to ``All''& 2\\\hline
    \cellcolor[HTML]{F0F0F0}\hypertarget{p11}{14}& \cellcolor[HTML]{F0F0F0}Adding an item first shows an empty state& \cellcolor[HTML]{F0F0F0}1\\\hline
  \end{tabular}
  \caption{Problems found in TodoMVC implementations}
  \label{tbl:Problems}
\end{table}

\Cref{tbl:Problems} describes in detail the specific faults that \Quickstrom exposed. The problem found in \texttt{angular-dart}, number 14, does not actually impede the overall operation of the application.
However, because it temporarily empties the list before re-populating it when adding a new item, this is a bug according to our formal specification. Because this is not explicitly forbidden by the official English specification, however, we consider it
a ``dubious'' case, and it could be considered ``correct'' by a more generous interpretation of the specification.
Of the remaining problems, three (Problems 1--3) are just unimplemented functionality or missing UI elements that are required by the specification. These problems appear
only in beta versions for the most part, and would likely be found in a cursory review.
The others all appear to be programming mistakes. Problems 4--6 are simple bugs that are easily found manually,  but the remaining problems (Problems 7--13) require nontrivial steps to uncover. In particular, problems often manifest when the user
does something unexpected after entering (but not committing) some input text, as in Problem 7 (the most common fault at four implementations) and Problem 8 (which also appeared in multiple implementations).
In addition, the interaction between filters and the ``Toggle all'' button is another common source of bugs, as in Problems 9--11.

Problem 11 is particularly involved to uncover, and could easily slip past cursory review. In order to reproduce this bug, the user must
create a to-do item, then immediately double click it to start editing, erase all text and press Enter (the item now appears deleted, but filters are still visible). Then, the user must
click the ``toggle all'' button, at which point the supposedly deleted item re-appears.

The bugs found in the various TodoMVC implementations run the gamut from trivial to complex. Notably, we found roughly as many faults in mature implementations as we did in beta ones.
We even found problems in all three of the ``Pure JavaScript'' examples (\texttt{vanillajs}, \texttt{vanilla-es6}, and \texttt{jquery}) considered as reference implementations by the TodoMVC specification.
These case studies demonstrate \Quickstrom's effectiveness as a bug-finding tool, even for mature software with extensive manual testing.

\subsection{Running Time and False Negative Rate}
\label{ssec:timing}

\begin{figure}

  \begin{tikzpicture}
  
  \begin{axis}[
    axis y line*=left,
    ymin=0, ymax=100,
    xlabel=temporal operator subscript (trace length),
    ylabel=false negative rate (\%),
  ]
  
  \addplot[smooth,tension=0.3,mark=o,cyan!50!black]
    coordinates{
      (1,100)
      (5,89.5)
      (10,64.5)
      (50,30.5)
      (100,21.4)
      (500,8.1)
  }; \label{plot_two}
  
  \end{axis}
  
  \begin{axis}[
    axis y line*=right,
    axis x line=none,
    ymin=0, ymax=220,
    ylabel=running time (s)
  ]
  \addlegendimage{/pgfplots/refstyle=plot_two}\addlegendentry{false negatives}
  \addlegendimage{/pgfplots/refstyle=plot_one}\addlegendentry{running time}
  \addplot[smooth,mark=x,red!70!black]
    coordinates{
      (1,12.849)
      (5,13.231)
      (10,14.432)
      (50,26.701)
      (100,41.833)
      (500,159.481)
  }; \label{plot_one}
  
  \end{axis}
  \end{tikzpicture}
  \caption{False negative rate and average running-time}
  \label{fig:graph}
\end{figure}
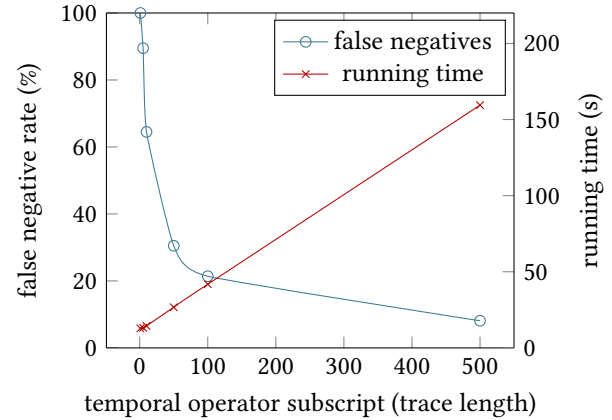

\Cref{fig:graph} summarises the relationship between the subscripts on temporal operators, test accuracy, and running time for our TodoMVC specification.
Specifically we measure the false negative rate (percentage of tests on faulty implementations that unexpectedly pass) for failing implementations and compare 
it to the average running time for testing passing implementations. This is because the TodoMVC specification consists only of safety properties: when checking a safety property, 
passing cases will always take more time than failing ones, as the testing tool will always exit as soon as a counterexample is found. Similarly, the only way that testing of safety properties 
could give inaccurate results is in the form of a false negative, because \Quickstrom will only report a test failure if a concrete counterexample is found. 
Conversely, when testing liveness properties, the situation would be reversed: failing cases would take the most time, and inaccurate results would be false positives.
For each subscript, each implementation was tested 10 times on a 2020 Apple MacBook Pro M1 with 16GB of RAM, although testing time is dominated by waiting for events, so performance of 
hardware does not greatly affect running time.

As can be seen, testing takes linearly more time but becomes logarithmically more accurate as the temporal subscript increases.  All of the faults found in TodoMVC can be exposed with a subscript of merely 50,
but the more involved faults such as Problem 11 are only found infrequently, resulting in flaky tests. The vast majority of faults 
are reliably found with a subscript of 100---the default value in \Quickstrom---and testing takes less than a minute (approx.\ 42 seconds for passing cases). After that, higher 
subscripts are still more likely to uncover faults, but there are diminishing returns in terms of faults found for time taken.

\section{Related and Future Work}

\subsection{Automated Browser Testing}
\label{ssec:targeted}
While tools such as Selenium WebDriver are now well established and enjoy widespread industry use, higher-level automation of
such acceptance testing is an area that has not yet been thoroughly explored. Like us, \citet{todomvc_learning} apply
their testing tool ALEX to the TodoMVC benchmark, however ALEX is based on learning-based testing without models
or specifications, and is therefore limited to finding inconsistencies between TodoMVC implementations. By contrast,
\Quickstrom generates tests based on user-provided logical specifications, and can therefore find more bugs, albeit with greater effort required for writing specifications.
We believe model inference techniques~\citep{lstar} such as those in ALEX and other model checking techniques such as counterexample-guided abstraction-refinement~\citep{cegar}
are highly compatible with \Quickstrom, and could potentially be used to make \Quickstrom more intelligently select actions and search for counterexamples---a kind of
\emph{targeted} property-based testing for LTL specifications~\citep{targeted,automating_targeted}.

\citet{layout} present a tool to automatically verify constraints on page \emph{layout} and appearance based on high-level specifications. While it is possible to verify some layout constraints with \Quickstrom, that is not
its primary purpose. \Quickstrom does not
presently feature any specific functionality to verify that pages are laid out correctly, focusing instead on behavioural specifications.
Thus, this tool is complementary with \Quickstrom.

\subsection{LTL for User Interfaces}
We are not the first to realise the suitability of LTL for describing user interfaces. \citet{ltltypesfrp} and \citet{jeltsch} simultaneously observed that, just as
logical formulae correspond to types of programs, LTL formulae can correspond to types for functional-reactive programs (FRP), including user interfaces and interactive applications. \citet{debuggingfrp}
used LTL formulae for testing and debugging of FRP programs.

The \emph{Model-View-Update} (MVU) architecture, pioneered by the Elm programming language~\citep{elm}, itself descended from FRP, is a simple design pattern for user interfaces that has now
become widespread, with variants for most programming languages and UI frameworks. At its core, it describes interactive applications with a type for the \emph{model} or application state $M$,
a type for the \emph{view} $V$, a type for \emph{actions} $A$, and a pair of functions $\mathit{display} : M \rightarrow V$ and $\mathit{update} : M \times A \rightarrow M$.
This model is highly compatible with the view of states and actions used in \Quickstrom.
As the \Quickstrom checker is not WebDriver-specific, we could repurpose it with custom executors to produce language- or framework-specific testing tools, allowing
\Specstrom and \QuickLTL specifications to be applied to these applications more directly.

\subsection{Other Executors and Debuggers} As neither \Specstrom nor \QuickLTL are specific to web applications, it is worth investigating other domains to see if they would be a
suitable fit. Other GUI frameworks such as GTK and Qt both have acceptance testing frameworks similar to Selenium WebDriver, and are obvious candidates, but there may be more interesting
use cases further afield: for example, our \Specstrom checker could also be attached to an emulator or a debugger for an embedded system, where actions and events take the form of IO signals and the
accessible state is the memory on the system. Model checkers based on LTL such as Spin are already used in the embedded systems area, so programmers in this area may already be amenable to LTL specifications.

\subsection{State Machine Specifications}
As previously mentioned, our specifications for both our egg timer example and our TodoMVC case study strongly resemble a specification of a state machine. Model-oriented property-based testing using state machine models was originally developed for
the implementation of QuickCheck for Erlang, which was used to find linearisable instances of race conditions~\citep{qcmutex}. The same version of QuickCheck was later used to test AUTOSAR implementations~\citep{autosar,autosar2}. This state machine idea has now been implemented for several other property-based testing systems,
including the original Haskell QuickCheck.
Unlike \Quickstrom, these frameworks require that the model capture the essential complexity of the application under test: it needs to be functionally complete to be a useful oracle. For a system that is conceptually simple,
such as a key-value database engine, this is not a problem, but for systems that are burdened with inherent complexity, such as a business application with many intricate rules, a useful model tends to grow as complex as the system itself.
\Quickstrom specifications can be more abstract: the engineer does not have to implement a complete functional model of their system, and is free to leave out details and specify only the most important aspects of their application.
For example, the timer specification given in \Cref{fig:timer} intentionally applies both to timers that reset when stopped and to timers that pause when stopped.

\subsection{\QuickLTL as a Temporal Logic}
While \QuickLTL is by definition a superset of other partial trace variants of LTL such as RV-LTL~\citep{rvltl}, we have not yet formally explored the relationship between \QuickLTL and conventional infinite-trace LTL dialects.
Recall that actions in \Quickstrom are divided into \emph{user actions}, under the control of the user, and \emph{events}, under the control of the application.
It is not reasonable to assume progress for all actions, as conventional LTL dialects do, as this would impose a requirement on applications that events must eventually occur if no user actions can be taken. The \emph{reactive LTL} of \citet{rvg_ltl} is designed specifically to address this
problem, and would serve as a good starting point for this investigation.

\subsection{Fault Injection} \citet{whyrandom} provide a theoretical justification for the surprising effectiveness of randomly testing
distributed systems with \emph{fault injection}---intentional simulation of network faults for testing purposes. This kind of fault injection is
often provided by tools such as Jepsen~\citep{jepsen} and \textsc{Elle}~\citep{elle}. While these tools are for systems like distributed databases, the same
fault injection technique may also be useful in \Quickstrom: Modern web applications often try to handle network interruptions gracefully, defaulting to local
storage or warning the user that the connection was lost. Simulating network faults would enable \Quickstrom specifications to test such scenarios.

\subsection{Security and Confidentiality} A drawback of our approach is that we can only write properties that are expressible in LTL, i.e.\ properties of a single trace.
While we can specify some properties that relate to security, such as our property requiring the user to log in to see the ``Finances'' page,
we cannot express security properties, such as information-flow security, in standard LTL as they are \emph{hyperproperties}---properties relating multiple traces~\citep{hyperproperties}.
While temporal logics exist to express hyperproperties~\citep{hyperltl,secltl,epistemicltl}, random testing for hyperproperties may not be as fruitful
as it is for \QuickLTL properties. As hyperproperties relate multiple traces,
a counterexample to a security property expressed as an $n$-hyperproperty would take the form of a $n$-tuple of traces, rather than a single trace.
This makes counterexamples to security properties significantly harder to find.
While \citet{qcifc2} report effectively testing security properties using randomised property-based testing, these tests were on abstracted models of security definitions
rather than on realistic systems. We expect that, when applied to real-world web applications with large amounts of state, counterexamples to security properties will not be easily
found by randomised exploration of the state space, and would likely benefit from the more \emph{targeted} approaches previously mentioned.

\subsection{Property-based Testing and Formal Methods} The specifications used for property-based testing resemble those used for formal verification of software.
In particular, QuickCheck test suites have served as sources of specifications for deductive verification of Haskell code~\citep{readysetverify}.
Our specifications too
resemble temporal logic specifications that one might find for formal tools such as TLA+, Spin or, most recently, Alloy 6.
In their work on
information flow, \citet{qcifc2} observe that
property-based testing is still valuable even in the context of formal verification, as it can eliminate the wasted effort of trying to prove a faulty or ill-specified system correct.
\citet{cogentcase} posit that property-based testing could be used as an incremental path towards more widespread adoption of formal verification among software engineers.
\Quickstrom very much fits into this theme, as specifications in \Specstrom could, with little modification, be transliterated for use in more formal, exhaustive tools.

\section{Conclusion}
We have presented \Quickstrom, a property-based browser testing framework for acceptance testing of web applications.
\Quickstrom users write formal specifications of their application's behaviour in our specification language \Specstrom,
based on our new dialect of Linear Temporal Logic for testing, \QuickLTL. With this specification, \Quickstrom will test
the application with hundreds of possible interactions, all generated automatically from the specification.

Our case studies demonstrate that \Quickstrom is an effective tool for finding non-trivial bugs in realistic web applications.
Writing a \Specstrom/\QuickLTL specification
enables programmers to find bugs more quickly and easily than by writing a comprehensive test suite with a browser testing framework.
But, more than that,
we hope that \Quickstrom will make formal specification and modelling, immensely powerful tools for improving software reliability,
more accessible to mainstream web application programmers.

\begin{acks}                            
 Many thanks to Rob van Glabbeek for his early feedback on the design of our logic.
\end{acks}

\bibliography{refs}



\end{document}